\documentstyle[preprint,aps]{revtex}
\begin{document}
\draft
\title{\bf Soliton excitations as emitted clusters on nuclear 
surfaces}
\author{R. A. Gherghescu $^{1,2}$, A. Ludu $^{1}$, and 
J. P. Draayer $^1$}
\address{$^1$ Department of Physics and Astronomy,
Louisiana State University, Baton Rouge LA 70803 USA}
\address{$^2$ National Institute of Physics and Nuclear
Engineering, P.O. Box MG-6, RO-76900 Bucharest-Magurele, Romania}

\maketitle

\begin{abstract}

This work reports on calculations of the deformation energy
of a nucleus plus an emitted cluster as a soliton moving on its surface. The
potential barrier against the emission of a soliton is calculated within
the macroscopic-microscopic method. The outer turning point 
of the barrier determines limitations on the geometrical and 
kinematical parameters for the formation of a surface soliton.
For large asymmetry, the two-center shell  model is used to assign a
structure to the soliton. Calculations for $^{248}$N\lowercase{o} with
the emission of a $^{40}$C\lowercase{a} soliton are reported; likewise
for  $^{224}$T\lowercase{h} with the emission of $^{16}$O.
Except for necked shapes at the very first stages of soliton
formation, the greatest portion of the deformation path displays
rather compact configurations with large neck radii. These shape
sequences correspond to allowable soliton velocities.
Close to and just beyond the touching point configuration, where the
shape becomes concave, the width and the velocity of the soliton
approach zero. The calculations suggest that the emission of a
$^{40}$C\lowercase{a} structure is quite probable due to a low potential
barrier, whereas the  emission of an $^{16}$O type soliton is rather
unlikely due to the higher penetration barrier. \\

PACS:05.45.Y, 21.60.Gx, 23.90.+w

\end{abstract}

\section{Introduction}

There are different theoretical models which describe cluster formation
and emission from nuclei, and most of them use nonlinear partial
differential equations.  A  fundamental understanding of  nonlinear terms in a
nuclear model reveals new phenomena and shapes more complicated than linear
theory suggests. In this paper we introduce a model for the cluster emission,
based  on soliton solutions of nonlinear equation.

Soliton structures have been analyzed within the
frameworks of hydrodynamics, nonlinear optics, solid state and plasma physics.
Experimental and theoretical results \cite{rebbi} suggest that solitons are
non-dispersive, localized waves executing uniform motion that can be described by
three interrelated parameters: amplitude $A$, half-width $L$, and velocity
$V$. Furthermore, these structures arise as analytical solutions of non-linear
dynamical systems, like the Korteweg de Vries (KdV) or Nonlinear
Schr\"odinger (NLS) equations.  Soliton theory has been applied many times in nuclear
physics, so far. For exemple, it provided very good descriptions of some localized
stable surface excitations \cite{fowler},  of spectroscopic factors in cluster and
$\alpha $ emission \cite{ludu}, and of quasimolecular spectra for
$\alpha$ plus heavy nuclei collisions, \cite{spectra}.
Also, the cubic and quintic NLS equations were used  in  three-dimensional models for
cluster emission  \cite{kartav}, providing results in very good agreement with  TDHF
simulations.

In order to look for the possibility of describing nuclear phenomena such as
cluster emission \cite{poe} by soliton formation on the nuclear surface,
it is necessary to assign a microscopic structure to the parent heavy
nucleus and the emitted soliton cluster. The microscopic substructure
further allows one to add shell corrections to the usual macroscopic
liquid drop energy and thus give a complete descripton of the system,
from the initial nucleus with no soliton substructure to one with a
soliton-like structure on its surface and on out to possible cluster
emission.

A straightforward way to accomplish this is to calculate shell effects
obtained from the single-particle levels of an asymmetric two-center
shell model. One of the centers is placed in the middle of a small
emitted sphere and the other is the center of the heavy fragment. This
approach allows a microscopic description of the nuclear evolution from
one to two independent quantum systems.

The procedure involves calculating the total potential energy as the sum
of the macroscopic energy and shell corrections which is then minimized,
which yields a barrier that increases as a function of the amplitude of
the soliton. Calculations have been performed for two possible reactions:

$^{248}$N\lowercase{o} $\rightarrow$ $^{208}$P\lowercase{b}
+$^{40}$C\lowercase{a} 

$^{224}$T\lowercase{h} $\rightarrow$ $^{208}$P\lowercase{b}
+$^{16}$O

\noindent 
We select spherical daughters and emitted clusters, in order to
fit the asymmetric two-center shell model we have constructed, which
goes from one sphere (parent) to two necked spheres and then to separation.

In the present model, we describe cluster emission processes by using such
soliton-like shapes on the nuclear surface of the heavy fragment. For a given cluster
geometry, we calculate the corresponding soliton parameters ($A$, $L$,
$V$) as functions of the separation parameter, that is along the static
path of the cluster emission process.

Solitary waves have been shown recently to exist on liquid drops, bubbles,
and shells \cite{ludu}. The non-linear hydrodynamic equations are related
with the KdV and mKdV equation generating localized patterns ranging from small
oscillations to nonlinear ones, including solitons. 
This model has a Hamiltonian structure, and such soliton-like excitations were
observed experimentally when the shape  oscillations of a droplet became nonlinear.
It is therefore natural to extend that this model to other drop-like systems, 
from neutron stars to hyperdeformed nuclei and fission.

In the first section we define the deformation space we that we work with 
and the shapes that can be obtained therein. In the second section we give
a short description of the  macroscopic-microscopic model we used in the
calculations. Emphazis has been given on the asymmetric two-center shell
model we constructed in order to approach solitonic shapes. Results for
the two reactions given above are presented in the third section.

\section{Space of deformation}

The problem of describing cluster emission (or nuclear fission) by a convenient
parametrization of the shape is not new, and it is a decisive factor determining
the amount of calculations. In the present model we use a new type of
parametrization, described by soliton shapes. As well as the other parametrizations
are in relation with the nuclear system and its quantization, the soliton
parametrization is related to the KdV and mKdV dynamics of the soliton.
Moreover, the soliton model takes profite from the high stability in time of such
shapes. While in general one associates solitons with macroscopic pictures,
it is nevertheless true that the soliton dynamics can be quantized
\cite{rebbi,fowler,spectra,birula} and the KdV (or mKdV) equations can be related to
quantum systems like those described by Schr\"odinger
equations, \cite{kartav,gridnev}, as we noted above.

Solitons on the surface are described by two asymmetrical spheres smoothly
joined one to another through a neck region (Fig. 1). There are three
independent geometrical parameters which form the space of deformation:
the distance $R$ between the centers of the two fragments, the emitted
small sphere radius $R_2$ and the neck sphere radius $R_3$. The neck
region is obtained by rolling a sphere of radius $R_3$  around the
symmetry axis. Such shapes are generated by the following equation
written in cylindrical coordinates:
\begin{equation}
\rho (z)= \left \{ \begin{array}{cc}
\sqrt{R_1^2-(z-z_1)^2}, & z_1 - R_1 \leq z \leq z_{c1} \\
\rho_3 - \sqrt{R_3^2-(z-z_3)^2}, & z_{c1} \leq z \leq z_{c2} \\
\sqrt{R_2^2 - (z-z_2)^2}, & z_{c2} \leq z \leq z_2+R_2
\end{array} \right.
\end{equation}
where the quantities not shown in Fig. 1 are: $z_3$--the
position of the center of the neck sphere on the symmetry axis, $z_{c1}$
and $z_{c2}$--the positions of the intersection planes of the two
fragements spheres with the neck sphere, and $z_1$ and $z_2$--the
positions of the two  spherical fragment centers. The heavy fragment
radius $R_1$ is obtained from total nuclear volume conservation. The
soliton solution along the $\theta $ direction
\begin{equation}
r_{surface}(\phi, \theta ,t)=
A(\phi ,t) \left [ sech \frac{\theta -Vt}{L} 
\right ]^2
\end{equation}
is characterized by the amplitude $ A $, or
the relative amplitude $a=A/R _{1}$, the half-width $L$, and the angular velocity
$V$. The soliton solutions  have a special shape--kinematic dependence,
$V \simeq A$ and $L \simeq 1/ \sqrt{A}$, that is, a higher soliton 
is narrower and travels faster \cite{rebbi}. This relation can be used to
experimentally distinguish solitons from other normal modes of excitations
(for example by calculating the reciprocal moment of inertia) \cite{ludu}.
The amplitude of the soliton is  related to the two-center
shell model by the relation  with $A=R-R_1 + R_2$. The halfwidth of the soliton is 
approximated with $2\rho (z_{c2})$, or with the diameter of the circular surface
within the separation plane between the emitted sphere and the rest of the shape
\begin{equation}
L=\frac{2R_2 A}{(1+a)(R_1-R_2)}.
\end{equation}
We mention that the cubic and quartic NLS equations are related to the mKdV equation
by a very simple exponential transform, and actually there is no essential difference
between the NLS and the mKdV solitons \cite{rebbi,gridnev}. As an example, we
notice the connection between the KdV equation and the nuclear potential in the
Schr\"odinger equation
\cite{gridnev}, or the relation between NLS solitons and coherent states
\cite{birula}. Some applications of the KdV or mKdV-solitons are macroscopic, but
the soliton solutions can be quantified by standard procedures
\cite{rebbi,spectra,rajaraman}.

\section{Macroscopic energy}

The deformation energy $E_{def}$
is calculated in a macroscopic-microscopic
approach:
\begin{equation}
E_{def}=E_{macro}+ \delta E_{shell}+ \delta P.
\end{equation}
The macroscopic part $E_{macro}$ includes the shape-dependent
components of the charged liquid drop:
\begin{equation}
E_{macro}=E_C+E_{Y+E},
\end{equation}
where $E_C$ is the Coulomb energy and $E_{Y+E}$ is the
surface or nuclear energy calculated within the
Yukawa-plus-exponential model \cite{scheid,kra}.

The Coulomb energy general form is the double-volume integral: 
\begin{equation}
E_C=\frac{1}{2}\int _V \int _V \frac{\rho _e (r_1)\rho _e(r_2)d^3r_1d^3r_2}
{r_{12}},
\end{equation}
which is split into four parts, two of them being equal to one another
\cite{poe}:
\begin{equation}
E_C=\frac{\rho^2_{1e}}{2}\int_{V_1}d^3r_1\int_{V_1}\frac{d^3r_2}{r_{12}}
+\rho_{1e}\rho_{2e}\int_{V_1}d^3r_1\int_{V_2}\frac{d^3r_2}{r_{12}}
+\frac{\rho^2_{2e}}{2}\int_{V_2}d^3r_1\int_{V_2}\frac{d^3r_2}{r_{12}}
\end{equation}
where $r_{12}=|r_1-r_2|$.
The first and the last terms represent the self-energies of the two
fragments, and the middle term is the Coulomb interaction between the
two fragments.

In cylindrical coordinates the three terms are given by:
\begin{eqnarray}
E_c&=& \frac {\rho _e ^2}{10} \int _{z'} ^{z''} dz
\int _{z'} ^{z''} dz_1 \int _0 ^{2\pi} d\varphi \int _0 ^{2\pi} 
d\varphi _1 \left ( \rho ^2 - \frac {z}{2} \frac {\partial \rho ^2}
{\partial z} \right ) \left [\rho ^2 _1 - \right. \nonumber \\
 & &\left. \rho \rho _1 \cos (\varphi - \varphi _1) + \rho \frac {\partial
\rho _1}{\partial \varphi _1} \sin (\varphi - \varphi _1) +
\frac{(z-z_1)}{2} \frac{\partial \rho ^2 _1}{\partial z_1} \right ]
[\rho ^2 +  \nonumber \\
 & &\rho _1 ^2 - 2\rho \rho _1 \cos (\varphi - \varphi _1) +
(z-z_1)^2 ]^{-1/2},
\end{eqnarray}
where $\rho = \rho (z, \varphi)$ is the nuclear surface equation, and
$z'$ and $z''$ are the intersections of the surface with $Oz$ axis.

The general form of the Yukawa-plus-exponential energy is:
\begin{equation}
E_{Y+E} =- \frac {a_2}{8\pi ^2 r_0^2 a^4} 
\int_V \int_V \left ( \frac
{r_{12}}{a}-2 \right ) \frac {\exp (- r_{12}/a)}{ r_{12}/a} 
d^3 r_1 d^3 r_2,
\end{equation}
where $r_{12}=|\bf r_1-r_2|$, $a$=0.68 fm accounts for the finite
range of nuclear forces, and $a_2=a_s(1-\kappa I^2)$. $\kappa $ is the
asymmetry energy constant, and the surface energy constant is $a_s$=21.13
MeV. In a similar way to the Coulomb part, one obtains three terms:
\begin{eqnarray}
E_{Y+E}&=&-\sum_{i=1}^{2}\frac{a_{2i}}{8\pi ^2 r_0^2a^4}
\int_{V_1}d^3r_1\int_{V_1}\left(\frac{r_{12}}{a}-2 \right )
\frac{\exp (-r_{12}/a)}{r_{12}/a}d^3r_2 \nonumber \\
& &-
\frac{2\sqrt{a_{21}a{22}}}{8\pi r_0^2a^4}
\int_{V_1}d^3r_1
\int_{V_2} \left(\frac{r_{12}}{a}-2 \right )
\frac{\exp{-r_{12}/a}}{r_{12}/a}d^3r_2
\end{eqnarray}
where $a_{2i}=a_s(1-KI_i^2)$, $I_i=(N_i-Z_i)/A_i$. 
For shapes with axial symmetry, each of these terms involving a
double-folded integration over the nuclear volume can be reduced
to a three-dimensional integral.

\section{The asymmetric two-center shell model}

The two-center shell model was developed by the Frankfurt school for
symmetric splitting \cite{scha} and for low asymmetry \cite{mar}. Here we
present the main steps of the two-center shell model we developed for
large asymmetry starting from another symmetrical two-center model
\cite{bad}.

The general Hamiltonian describing the evolution of the level scheme of
the two-center shell model is based on two oscillators which split from
an initial common oscillator.
The usual spin-orbit interaction and the $\bf l^2$ term are constructed as
depending on the mass asymmetry. Thus, the total Hamiltonian reads:
\begin{equation}
H=H_{osc}+V({\bf l \cdot s})+V({\bf l^2})
\end{equation} 
where $H_{osc}$ is the two-oscillator Hamiltonian, and $V({\bf l \cdot s})$ 
and $V({\bf l^2})$ are the spin-orbit and the ${\bf l^2}$ potentials.

\subsection{The diagonalization basis}

The oscillator part of the Hamiltonian, $H_{osc}$, is given in
cylindrical coordinates by:
\begin{equation}
H_{osc}=-\frac{\hbar^2}{2m_0}
\left [\frac{\partial ^2}{\partial \rho ^2}+\frac{1}{\rho}
\frac{\partial}{\partial \rho}+\frac{1}{\rho ^2}
\frac{\partial ^2}{\partial \phi ^2}+
\frac{\partial ^2}{\partial z^2} \right ]+V(\rho ,z),
\end{equation}
where the asymmetric two-center oscillator potential has the form:
\begin{equation}
V(\rho ,z)=\frac{1}{2}m_0 \left \{ \begin{array}{clc}
\omega ^2_{\rho _1}\rho ^2+\omega ^2_{z_1}(z+z_1)^2 &,& z<z_0
\nonumber \\
\omega ^2_{\rho _2}\rho^2+\omega ^2_{z_2}(z-z_2)^2 &,& z\geq z_0 .
\end{array} \right.
\end{equation}
Here $z_0$ is the separation plane coordinate between the two asymmetric
systems on the symmetry axis Oz. Since we consider the case of two
asymmetric spheres, $\omega _{\rho _1}=\omega _{z_1}=
\omega _1$ and $\omega _{\rho _2} =\omega _{z_2}=\omega _2$.

To obtain an appropiate basis for this system, we consider the
intermediate case when $\omega _{\rho _1}=\omega _{\rho_2}=\omega _1$.
Then the potential is:
\begin{equation}
V(\rho ,z)=\frac{1}{2}m_0 \left \{ \begin{array}{clc}
\omega ^2_1\rho ^2+\omega ^2_1(z+z_1)^2 &,& z<0
\nonumber \\
\omega ^2_1\rho^2+\omega ^2_2(z-z_2)^2 &,& z\geq 0 .
\end{array} \right.
\end{equation}
At this point the intermediate Hamiltonian is separable, and 
one gets the eigenfunctions \cite{bad}:
\begin{equation}
\Phi _m(\phi)=\frac{1}{\sqrt{2\pi}}\exp{(im \phi)} 
\end{equation}
for the axial degree of freedom and 
\begin{equation}
R_{n_{\rho}|m|}(\rho)=\sqrt{\frac{2\Gamma (n_{\rho}+1)
\alpha _1^2}{\Gamma (n_{\rho}+|m|+1)}}
\exp{\left (-\frac{\alpha _1^2\rho ^2}{2}\right )}
(\alpha _1^2\rho ^2)^{\frac{|m|}{2}}
L_{n_{\rho}}^{|m|}(\alpha _1^2\rho ^2)
\end{equation}
for describing radial motion where $\alpha _i =(m \omega _i/ \hbar )$,
$\Gamma(x)$ is the gamma function, and $L_{n_{\rho}}^{|m|}(x)$ is the
Laguerre polynomial. As one knows, the oscillator energy for oscillations
in the plane perpendicular on the symmetry axis is:
\begin{equation}
E_{\rho , \phi }=\hbar \omega _{\rho}(2n_{\rho}+|m|+1)
\end{equation}
where $\omega _{\rho}=\omega _1$.

Solving the third equation which accounts for oscillations along the
symmetry axis, we have different solutions for the two regions
of the nuclear shape. According to the $z$-dependent potential,
\begin{equation}
V(z)=\frac{1}{2}m_0\left \{ \begin{array}{clc}
\omega _1^2(z+z_1)^2 &,& z<0 \nonumber \\
\omega _2^2(z-z_2)^2 &,&z \geq 0 
\end{array}
\right.
\end{equation}
where $z_1$ and $z_2$ are the centers of the heavy and light 
spherical fragments, respectively, one obtains the
differential equations
\begin{eqnarray}
\left [\frac{d^2}{dz^2}+\frac{2m_0E_z}{\hbar ^2}
-\frac{m_0^2\omega _1^2(z+z_1)^2}{\hbar ^2} \right ]
Z_{\nu _1}(z)=0  & &\mbox{$, z<0$} \\
\left [\frac{d^2}{dz^2}+\frac{2m_0E_z}{\hbar ^2}
-\frac{m_0^2\omega _2^2(z-z_2)^2}{\hbar ^2} \right ]
Z_{\nu _2}(z)=0 & &\mbox{, z$\geq$ 0 }.
\end{eqnarray}
At this point it is important to mention that the $z=0$
plane is the intersection plane between the two systems
with $\omega _{\rho _1}=\omega _{\rho _2}$, whereas the
``real'' intersection between the asymmetric spherical systems
is at $z=z_0$. The solution for the $z$-dependent Hamilton
equation will be:
\begin{equation}
Z_{\nu}(z)=\left \{ \begin{array}{clc}
C_{\nu _1}\exp{\left [ -\frac{\alpha_1 ^2(z+z_1)^2}{2}\right ]}
H_{\nu _1}[-\alpha _1(z+z_1)] &,&z<0 \nonumber \\
C_{\nu _2}\exp{\left [ -\frac{\alpha _2^2(z-z_2)^2}{2}\right ]}
H_{\nu _2}[\alpha _2(z-z_2)] &,& z\geq 0
\end{array}\right .
\end{equation}
where $C_{\nu _1}$ and $C_{\nu _2}$ are normalization constants,
and $H_{\nu}(z)$ are the Hermite functions.

As can be seen from these results, four quantities need to be determined:
the two quantum numbers $\nu _1$ and $\nu _2$ and the two normalization
constants $C_{\nu _1}$ and $C_{\nu _2}$.
These quantities can be calculated from a system of four equations.
From the normalization condition 
\begin{equation}
\int_{- \infty}^{\infty}|Z_{\nu}(z)|^2dz=1,
\end{equation}
from the continuity of the $z$-wave function and its derivative
at $z$=0
\begin{equation}
Z_{\nu _1}(z=0)=Z_{\nu _2}(z=0),
\end{equation}
\begin{equation}
Z^{\prime}_{\nu _1}(z=0)=Z^{\prime }_{\nu _2}(z=0),
\end{equation}
and from the energy matching condidtion along the O$_z$ axis
\begin{equation}
\hbar \omega _1(\nu _1+0.5)=\hbar \omega _2(\nu _2+0.5).
\end{equation}
From these, a basis for diagonalization of the potential differences to
obtain the real energy values can be calculated.

\subsection{The asymmetric oscillator system}

Once we have total wave functions, we have to determine
differences between the diagonal Hamiltonian and the real one.
First, the oscillator Hamiltonian has to provide the initial
oscillator potential when there is only one heavy sphere
(starting point). For this initial configuration the difference that
needs to be  diagonalized is
\begin{equation}
\Delta V^{sphere}(z)=\left \{ \begin{array}{clc}
\frac{1}{2}m_0[\omega _1^2(z+z_1)^2-\omega _2^2(z-z_2)^2]
 &,& z\geq 0 \nonumber \\
0 &,& z<0 .
\end{array}
\right .
\end{equation}
For the next stages of deformation, the difference between 
the $z$-dependent oscillator potentials that needs to be diagonalized is:
\begin{equation}
\Delta V(z)=\left \{ \begin{array}{clc}
0 &,& z<0 \nonumber \\
\frac{1}{2}m_0[\omega _1^2(z+z_1)^2-\omega _2^2(z-z_2)^2
&,& 0\leq z \leq z_0 \nonumber \\
0 &,& z>z_0 .
\end{array}
\right .
\end{equation}
As for the difference in the $\rho$-dependent oscillator potential, this
only exists for intersecting spheres and is given by:
\begin{equation}
\Delta V(\rho)= \left \{ \begin{array}{clc}
0 &, & z \leq z_0 \nonumber \\
\frac{1}{2}m_0(\omega ^2_1-\omega ^2_2)\rho ^2 & ,&  z > z_0 
\end{array}
\right .
\end{equation}
or, if written as an operator, the quantity to be diagonalized is given
by:
\begin{equation}
\Delta V(\rho)=\frac{1}{2}m_0(\omega ^2_1-\omega ^2_2)\rho ^2
\Theta (z-z_0)
\end{equation}
where $\Theta(z)$ is the Heaviside function. The difference
$\Delta V(\rho)$ is zero for the initial spherical configuration.

Once $\Delta V(z)$ and $\Delta V(\rho)$ are diagonalized and added
to the oscillator energy of the sphere + ellipsoid system, which is
\begin{equation}
E=\hbar \omega _1 [2n_{\rho}+|m|+\nu _1+1.5],
\end{equation}
the level schemes of the two intersecting asymmetric oscillators with
frequencies $\omega _1$ and $\omega _2$ are obtained.

\subsection{Spin-orbit and orbit-orbit interactions}

The spin-orbit ($\bf l\cdot s$) and orbit-orbit (${\bf l^2}$) interaction
terms generate the necessary single-particle level splitting to obtain the
correct schemes of the individual fragments after separation.

The use of deformation dependent form of these operators has been
introduced in \cite{rij} for the Nilsson model and in \cite{mar} for
the two center shell model; instead of the ${\bf l}$ operator one
introduces:
\begin{equation}
{\bf l}=\frac{\nabla V \times {\bf p}}{m_0 \omega ^2}
\end{equation}
where $V$ is the asymmetric two-center oscillator potential.
The usual expression for the two operators are:
\begin{eqnarray}
V({\bf l \cdot s}) &=&-2\kappa \hbar \omega {\bf l \cdot s} \nonumber \\
V({\bf l^2}) &=&-\kappa \mu \hbar \omega {\bf l^2}.
\end{eqnarray}
Since one obtains the level schemes of two nuclei 
which lie in different mass regions, the strength parameters
of the interactions $\kappa$ and $\mu $ will be different.
The values we use for these parameters are:
\begin{equation}
\begin{array}{ccc}
\kappa _n=0.0588 & &\kappa _p=0.0592 \nonumber \\
\mu _n=0.328 & &\mu _p=0.335 \nonumber 
\end{array}
\end{equation}
for actinide region, and for light nuclei region:
\begin{equation}
\begin{array}{ccc}
\kappa _n &= &\kappa _p=0.0601 \nonumber \\
\mu _n &= &\mu _p=0.448. \nonumber 
\end{array}
\end{equation}
Since the strength parameters are different for the asymmetric
regions of the nuclear shape, they become $z$-dependent
operators as follows:
\begin{equation}
\begin{array}{ccc}
\kappa \cdot \hbar \omega (z)=\kappa _1 \cdot \hbar \omega _1
+(\kappa _2 \cdot \hbar \omega _2 -\kappa _1 \cdot
\hbar \omega _1)\Theta (z-z_0) && \\
\kappa \mu \cdot \hbar \omega (z)=
\kappa _1 \mu _1 \cdot \hbar \omega _1+
(\kappa _2 \mu _2 \cdot \hbar \omega _2-
\kappa _1 \mu _1 \cdot \hbar \omega _1)\Theta (z-z_0).
\end{array}
\end{equation}
To obtain a Hermitian operator for $V({\bf l \cdot s})$ and
$V({\bf l^2})$ one has to use the anticommutator \cite{mar}:
\begin{equation}
\begin{array}{ccc}
V({\bf l \cdot s}) = -\left [\kappa \cdot \hbar \omega (z),
\frac{\nabla V\times {\bf p}}{m_0 \omega ^2}{\bf s} \right ] , &&\\
V({\bf l^2}) = -\frac{1}{2}\left [\kappa \mu \cdot \hbar \omega (z),
\left (\frac{\nabla V \times {\bf p}}{m_0 \omega ^2 }\right)^2 \right ].
\end{array}
\end{equation}
For the dependence of $\kappa $ and $\mu $ with respect to the elongation, we choose
a linear dependence for the oscillator frequency along the $z$-axis:
\begin{equation}
\kappa _i=\kappa _0+\frac{\omega _i-\omega _0}{\omega _{if}-\omega _0}
(\kappa_{if}-\kappa _0)
\end{equation}
and the same law of variation for $\mu$. Here $i$=1,2, $\kappa _0$
is the value for the initial nucleus and $\kappa _{if}$ for the 
final one.

Finally one has to diagonalize the potential:
\begin{equation}
\Delta V(\rho ,z)=\Delta V(z)+\Delta V(\rho)+V({\bf l \cdot s})+V({\bf l^2})
\end{equation}
together with the diagonal term of the two-center oscillator potential.
The model provides the evolution from an initial 
level scheme toward two asymptotically independent single-particle schemes. With the
introduction of large asymmetry between fragments, the shapes can simulate the
existence of a soliton on the nuclear surface and assign it a microscopic
structure.

\section{Shell corrections}

The level scheme of a soliton shape is used to obtain the shell
corrections of the system. As the soliton is assimilated with an
emerging fragment, it will provide the shell correction
value of the independent nucleus of similar shape.
Shell corrections are obtained by means of the Strutinsky procedure
\cite{str}. One defines the shell correction energy as the difference
between the total sum of the energy levels and a smoothed part
of the spectrum:
\begin{equation}
\delta E=\sum _{\nu}2E_{\nu}-\tilde{U}.
\end{equation}
One calculates the smoothed part $\tilde{U}$ with the help of
a smoothed level density function $\tilde{g}(\epsilon)$, which is
obtained by averaging the real distribution $g(\epsilon)$ over the
whole energy spectrum:
\begin{eqnarray}
  \tilde{g}(\epsilon) & = & \frac{1}{\gamma}\int _{- \infty} ^{\infty}
  \zeta \left (\frac{\epsilon - \epsilon ^{\prime}}{\gamma} \right )
  g(\epsilon ^{\prime})d \epsilon ^{\prime} \nonumber \\
  & = & \frac{1}{\gamma}\sum _{i=1} ^{\infty} \zeta \left (
  \frac{\epsilon -\epsilon ^{\prime}}{\gamma} \right ),
  \end{eqnarray}
where $\gamma =\Gamma/\hbar \omega$ and $\zeta(x)$ is the
smoothing function. A common smoothing function is provided by
\begin{equation}
  \zeta (x)=\frac{1}{\sqrt{\pi }}e^{-x^2}\sum _{k=0} ^m a_{2k}
  H_{2k}(x),
  \end{equation}
  where $H_n(x)$ are the Hermite polynomials. The coefficients $a_{2k}$
are
\begin{equation}
  a_{2k}=\frac{H_{2k}(0)}{2^{2k}\cdot (2k)!} .
\end{equation}
The Fermi energy $\tilde{\lambda}$ of the smoothed level distribution
  is calculated as a solution of the particle number
  conservation:
\begin{equation}
  N_p=2\int _{-\infty}^{\tilde {\lambda}}\tilde {g}(\epsilon)d\epsilon .
  \end{equation}
Then, the total energy of the uniform level distribution $\tilde{U}$,
  reproducing the microscopic part which is not subjected to local
  fluctuations of the spectrum, is obtained as:
\begin{equation}
  \tilde{U}=2\hbar \omega \int _{-\infty}^{\tilde {\lambda}}
  \tilde{g}(\epsilon)\epsilon d\epsilon.
  \end{equation}
After performing the calculations, one obtains the following
  formula, which can be used directly
\begin{eqnarray}
  \delta U & = & \sum _{\nu}
  \{ \epsilon _{\nu} \lbrack 1-erf(x_{F_{\nu}}) \rbrack \nonumber \\
 & & \mbox{} +
  \frac{e^{-x_{F_{\nu}} ^2}}{\sqrt{\pi}} \cdot
  \lbrack 2\epsilon _{\nu}\sum _{k=1} ^m a_{2k}H_{2k-1}+
  \tilde{\gamma}a_{2m}H_{2m} \rbrack \},
  \end{eqnarray}
where $erf(x)$ is the error function. 
Usually one chooses the upper order of the Hermite polynomials to be $m$=3.
The variable $x_{F_{\nu}}$ is given by
\begin{equation}
x_{F_{\nu}}=\frac{\epsilon _{\nu}-\epsilon _{F}}{\gamma },
\end{equation}
Shell corrections are calculated
separately for protons and neutrons, and the results are added.

\section{Results}

A first look at the energetic behavior of an emerging soliton
is given in Fig. 2 in terms of the macroscopic energy surfaces. The
LHS macroscopic energy surface corresponds to the formation of 
$^{40}$C\lowercase {a} on the surface of $^{248}$N\lowercase{o}, whereas
the RHS represents the $^{16}$O-like soliton on the surface
of $^{224}$T\lowercase{h}. Variation along the  elongation $R$
corresponds to the increment in the soliton amplitude along the symmetry
axis. A larger neck radius $R_3$ corresponds to a larger half-width
$L$. The rear plane at $R$=0 is the spherical state of the
system. Then the energy increases monotonously with a higher slope for
small values of the neck radius. As $R_3$ increases, the energy increase is smoothed
by the necking. With the enhancement of the kinetic energy of the
soliton the half-width becomes larger,  except in the first stages of the process
where the neck radius is very small. The ridge in energy
has a maximum at the near touching spheres configuration for both
reactions. The slope continues to increase for large $R_3$, beyond the touching
point value of the  elongation $R$.

The addition of shell corrections yields the total deformation energy
shown in Fig. 3.  As a first observation note the pronounced
deformed ground state of $^{248}$N\lowercase{o} as the first minimum in
energy moves to $R>$0, and a much less but still deformed ground state for
$^{224}$T\lowercase{h}.  For both emerging solitons it is obvious that the energy
path corresponds to large half-width values up to the top
of the energy ridge; then they abruptly turn towards rupture 
point shapes ($R_3$=0). Hence, these potential energy surfaces suggest a 
three-dimensional curve as the path of minimum energy in the cluster emission.
The potential barrier formed along the path of minimum energy values
is obtained by minimization of the total energy in the multi-dimensional
deformation space.

The static paths, which a soliton with the internal structure of an
emitted cluster has to follow, have been plotted on the contour maps
of the energy surface in ($R$,$R_3$) coordinate space in Fig. 4.
Again the LHS plot is the $^{40}$C\lowercase{a} emission, and the RHS one
corresponds to $^{16}$O. Apart from the first $R$ values, where $R_3$
is small, the solitons bypass the first energy peak by taking 
large neck radius values, i.e. large half-widths.
 As the energy increases on the large $R_3$ side,
the static path for both cases changes direction 
reaching the scission point where the clusters are emitted.

The two barriers are plotted in Fig. 5 with a full line, together with
the macroscopic energy (dotted line) and the shell corrections
(dashed line). One can see how the deformed ground state of
$^{248}$N\lowercase{o} is formed (LHS plot): due to
shell corrections, the first minimum is at about 6.8 MeV of the total
energy. This point becomes the ground state and the whole barrier
in front of the emerging soliton is shifted with respect to this
value. A two-humped barrier no higher than about 1.2 MeV blocks the
$^{40}$C\lowercase{a} emission.

The situation is different for the emission of $^{16}$O from
$^{224}$T\lowercase{h}. The ground state is only
slightly deformed. A rather high one-hump barrier of about 11 MeV
extends along the whole range of elongation $R$ up to the scission
point. Shell corrections decrease slightly the macroscopic
energy values. The decrease is mainly due to the double-magic character of
$^{208}$P\lowercase{b} which forms as the cluster emerges.

One can state that $^{40}$C\lowercase{a}-like solitons are energetically
favored to form on the nuclear surface of a very heavy nucleus
as $^{248}$N\lowercase{o}. The formation of $^{16}$O-like soliton
on $^{224}$T\lowercase{h} is not energetically favored due to the
high and large potential barrier it has to penetrate.

The relative velocity distribution $V$
of the two presumed solitons along the minimum
energy path, together with the scaled values of the 
halfwidth $L$ and the relative amplitude $a=A/R_1 $,
are plotted in Figs. 6.  We investigated the evolution of these soliton
parameters (as defined in section I), which are a function of the static energy 
evolution, parametrized by the distance between centers $R$. 

In the first stages, the tendency is that the amplitude and
half-width increase with the elongation parameter, when the emitted cluster is
emerging out from the parent nucleus  (since their non-overlaping sector is
increasing). During the  formation of the cluster the half-width  remains practically
constant, since the surface energy controls this stage. When the two nuclei are well
separated, the soliton envelope hardly fits the two spheres, and in this limit, the
half-width approaches zero value. This gives the limiting configuration for this
soliton model. These values of the half-width
$L$ (solid line) are compared with those obtained analyticaly in \cite{ludu} directly
from the soliton amplitude, within the frame of the nonlinear liquid drop model 
($L$-dashed line). We notice a good agreement for the half-widths  within
the range $R\simeq 4.5 - 14$ Fm. The hydrodynamic soliton model is not
valid anymore for separation parameter $R$ smaller than 4-5 Fm, because of the
dominating shell effects in this range. This can be noticed in a comparison
between Fig. 2 and Fig. 3 for $R\leq 4-5$ Fm.
For the first $R$ values, the static paths follow the first energy peak, and
jump from small toward large values for $R_3$, Fig. 4, providing small
half-widths (Figs. 6, $L$-solid line), while a pure hydrodynamic soliton 
would have larger half-widths for this range ($L$-dashed line).
In the above range of validity of the soliton model, we calculate the relative
velocity of the soliton ($V$-dashed line), \cite{ludu}. The velocity is
increasing with the amplitude of the soliton, hence with the elongation of the
cluster-like emission shape. Fig. 6a displays the $^{40}$C\lowercase{a}
emission, and Fig. 6b represents the $^{16}$O emission.
Soliton shapes at the begining and the end of the
process are also shown. 
For lighter nuclei (like $^{16}$O) the evolution of the parameters is
smooth and monotonic. In the case of heavier nuclei ($^{40}$C\lowercase{a})
we obtained some oscillations in width and velocity, during
the first half of the emission process, which can be related with the
oscillations produced by the shell effects in the $R_3$ parameter.

\vskip 1cm

This work was supported by the U.S. National Science Foundation through a regular
grant, No. 9970769, and a Cooperative Agreement, No. EPS-9720652, that includes
matching from the Louisiana Board of Regents Support Fund. RAG is grateful
to Prof. J. P. Draayer for a postdoctoral appointment in the Department
of Physics and Astronomy, Louisiana State University, Baton Rouge. 

\vfill
\eject

\newpage

Figure captions
\vskip 0.2cm

Fig.1
Deformation space for two-center shell-model calculations.
The neck radius $R_3$ can vary from zero (intersecting spheres) to
infinity (compact shapes). $R$ (the distance between the 
two centers) and $R_2$ (radius of the emitted
fragment) are also independent coordinates.

\vskip 0.15cm
Fig.2 
Macroscopic potential energy surfaces for $^{40}$C\lowercase{a}
emission from $^{248}$N\lowercase{o} (left-hand-side plot)
and for $^{16}$O emission from $^{224}$T\lowercase{h} (right-hand-side
plot), as function  of elongation $R$ and neck radius $R_3$. First
maximum appears close to the touching point configuration in both cases
($R_3$=0).

\vskip 0.15cm
Fig.3
Total potential energy surfaces (macroscopic plus shell corrections)
for $^{40}$C\lowercase{a} emission from $^{248}$N\lowercase{o}
(left-hand-side plot) and $^{16}$O from $^{224}$T\lowercase{h}
(right-hand-side plot). The deformed ground state of
$^{248}$N\lowercase{o} is revealed as a minimum along $R_3$ axis at its
origin. For both surfaces the closest energy maximum occurs at the tangent
sphere configuration. The maximum energy value for larger $R_3$ is not
reached in the figure.

\vskip 0.15cm
Fig. 4
Contour plot of Fig. 3 with static path (dashed line)
for $^{40}$C\lowercase{a} emission (left-hand-side plot) and
$^{16}$O emission (right-hand-side plot). As the elongation increases
the amplitude of the soliton increases together with the half-width which
is proportional to the neck radius $R_3$. Once the touching point
maximum is bypassed, both shapes decrease rather abruptly through necking
towards scission.

\vskip 0.15cm
Fig. 5
The barriers along the static path for $^{40}$C\lowercase{a}
emission (left-hand-side plot) and $^{16}$ (right-hand-side plot),
together with macroscopic energy (dotted lines) and shell corrections
(dashed lines). Shell corrections increase the total energy of the first
energy minimum for $^{248}$N\lowercase{o}, thus the two-humped
barrier is not higher than about 1.2 MeV. $^{16}$O emission from
$^{224}$T\lowercase{h} has a barrier of about 11 MeV. 

\vskip 0.15cm
Figs. 6 
The evolution of $a=A/R_1$, $L$ (with shell corrections solid line, and
without shell corrections dashed line) and
$V$ parameters in relative units, versus the elongation $R$ in fm. The
corresponding nuclear configurations are plotted for two situations: for the
initial stage when the emitted cluster is only slightly
displaced off the common center, and the final stage when the 
two nuclei are almost separated. Fig. 6a
displays the $^{40}$C\lowercase{a} emission and Fig. 6b
the $^{16}$O emission. In the later case, oscillations can be seen
in the soliton parameters related to the shell corrections.

\end{document}